# All in one: holographic microscopy unveils bacterial transport from single cells to population

Lucie Klopffer[a,b], Simon Becker[b], Laurence Mathieu[c], Nicolas Louvet[b*],

[a] Université de Lorraine, CNRS, LCPME, F-54000 Nancy, France ; [b] Université de Lorraine, CNRS, LEMTA, F-54000 Nancy, France ; [c] EPHE, PSL, UMR CNRS 7564, LCPME, F-54000 Nancy, France

*Corresponding authors :

Nicolas Louvet ; LEMTA, 2 Avenue de la Forêt de Haye, BP 90161, F-54505 Vandoeuvre-lès-Nancy, France **;** nicolas.louvet@univ-lorraine.fr

## Abstract

Prior to pioneer surface adhesion, bacteria have to navigate in flows, often in confined environments. While much is known about their individual swimming dynamics, our understanding of their transport properties at the population level remains limited. This is primarily due to the experimental challenges associated with tracking, in three dimensions and under flow, a large sample of these microorganisms. Here we investigate, through fast digital holographic microscopy (DHM), a suspension of *Shewanella oneidensis* MR-1 bacteria in a confined Poiseuille flow. Based on the analysis of several thousand Lagrangian trajectories, we first demonstrate the ability of DHM to discriminate between the fraction of bacteria that adhere to walls, those that are only convected by the flow, and those for which motility helps them migrate through the channel and cross streamlines. Focusing on the motile part of the population, we report new experimental results concerning the spatial and orientational distributions across the confinement direction. In the central part of the flow, migration and shear trapping are observed and well described by existing kinetic models. Close to the walls, averaging over the entire motile population clearly shows an accumulation layer but with no net orientational order, which contradicts models that neglect hydrodynamic interactions with solid surfaces. We think that DHM coupled with the analysis of large-scale bacterial populations will help to understand their transfer mechanisms from the bulk to surfaces.

## Introduction

Understanding the transport properties of planktonic bacteria in flows and their transfer to surfaces is of significant interest as it is the precursor step to biofilm formation (Berne *et al*., 2018; Busscher and Van Der Mei, 2006; Klopffer *et al*., 2024). It influences many natural or industrial processes (Donlan *et al*. 2016; Satpathy *et al*., 2016; Sentenac *et al*., 2022) as in healthcare (Koo *et al*., 2017) or even in spacecraft (Diaz *et al*., 2019; Goeres *et al*., 2023) and is of prime importance for antibiofilm materials strategies (Zou *et al*., 2025). In these various applications, while passive transfer mechanisms like diffusion (due to Brownian motion), sedimentation (density differences between bacteria and surrounding fluid) and fluid advection contribute to the transfer from bulk to surfaces (Busscher and Van Der Mei, 2006; Carniello *et al.*, 2018; Palmer *et al.*, 2007), the role of bacteria motility and its coupling with a flow is of first importance (Clement *et al.*, 2016; Lauga, 2016; Wadhwa and Berg, 2022). For instance, during the last decades, numerous studies highlighted complex near-surface swimming behaviors like bacterial accumulation near walls (Berke *et al.*, 2008; Clement *et al.*, 2016; Drescher *et al.*, 2011; Li *et al.*, 2008a), reduced swimming speeds when approaching surfaces (Elius *et al.*, 2023; Frymier *et al.*, 1995; Lauga and Powers, 2009; Qi *et al.*, 2017), body reorientation *(Bianchi et al., 2017; Lauga, 2016; Molaei et al.*, 2014), circular trajectories close to walls (DiLuzio *et al.*, 2005; Khong *et al.*, 2021; Lauga, 2016; Li *et al.*, 2008a), upstream swimming against flow (Hill *et al*., 2007; Kaya and Koser, 2012; Mathijssen *et al*., 2019) or rheotaxis (Jing *et al*., 2020; Kaya and Koser, 2012; Marcos *et al*., 2012; Morales-Soto *et al*., 2015). Far from surfaces, bacteria experience the flow field leading to active Bretherton–Jeffery orbits (Junot *et al.*, 2019) and complex concentration distribution profiles, depending on the swimming behavior of the bacteria and the flow conditions. Yet, aside from a limited number of experimental studies (Darnige *et al.*, 2017; Figueroa-Morales *et al.*, 2020; Junot *et al.*, 2019; Mathijssen *et al.*, 2019; Molaei *et al.*, 2014; Molaei and Sheng, 2016; Rusconi *et al.*, 2014; Torres Maldonado *et al.*, 2024), there is still a shortage of quantitative and statistically robust experimental data on bacterial motion in the bulk, especially the full time resolved three-dimensional bacteria dynamics, which hampers the validation and refinement of theoretical models (Elgeti and Gompper, 2013; Ezhilan and Saintillan, 2015; Ganesh *et al.*, 2023; Junot *et al.*, 2019).

One major challenge lies in accurately detecting and tracking bacterial trajectories in the bulk under flow. These difficulties arise from both the fluid dynamics and the intrinsic bacterial properties of bacteria, such as their small size and their ability to cross streamlines in the case of motile populations (Acres *et al.*, 2021; Dubay *et al.*, 2023; Taute *et al.*, 2015). Most experimental studies use conventional microscopy techniques such as phase contrast or fluorescence. While these approaches are simple to set up, they offer limited depth of field, which makes them impossible to resolve the full three-dimensional bacteria trajectories over a long period of time (Palma *et al.*, 2022). Only a small number of three-dimensional techniques have emerged to overcome this problem. Piezzo-motor objective and motorized stage coupled to real-time imaging was developed to record 3D trajectories with high resolution (Darnige *et al.*, 2017). Although this setup is very sensitive, it only allows individual bacteria to be tracked, thus losing the concept of population tracking. Digital holographic microscopy (DHM) is a powerful, time-resolved, 3D technique for tracking complex bacterial suspensions in flow. More than hundreds of bacteria can be tracked simultaneously over a long period of time (minutes), enabling an analysis at the individual or population scales and providing a high level of statistical accuracy. However, some drawbacks limit the use of DHM. Processing holograms with high bacterial densities is extremely time-consuming and often necessitates access to powerful computing clusters (Molaei and Sheng, 2016). Moreover, the low intrinsic contrast between bacteria and the background often requires advanced and optimized hologram processing. This significant computational demand hinders the adoption of the technique.

In this paper, we implement a Gabor in-line digital holographic microscope to access the 3D Lagrangian bacterial trajectories of a *Shewanella oneidensis* MR-1 suspension. Our first contribution comes from our DHM algorithm which enables the analysis of thousands of holograms while requiring minimal computational resources and so a fast and easy implementation in any laboratory. Also, while most experiments in the literature are conducted in a very diluted regime, our experimental facility allows higher working concentration that enables the analysis of thousands of bacteria providing stronger statistical significance. Eventually, performing analysis at the individual and population scales, we report new results on *Shewanella oneidensis* strain that is not characterized in terms of motility and dynamic under flow. This monotrichous bacteria demonstrates fast and persistent swimming that drives to a rarely observation of low shear trapping and unexpected super active Jeffery loops even at high shear rates.

## Materials and Methods

### *Shewanella oneidensis* strains and inoculum preparation

Experiments were performed using *Shewanella oneidensis* MR-1 WT as a model strain (Teal *et al.*, 2006). A Δfla deletion mutant of *Shewanella oneidensis* MR-1 (Armitano *et al.*, 2013), was used as a control. Motility tests were performed and confirmed the absence of motility of the mutant (Fig. S1). Prior to assay, the requisite strain of *S. oneidensis* was recovered from a glycerol stock at -80°C onto Lysogeny Broth agar plates. After 2 days incubation at 30°C, several isolated colonies of *S. oneidensis* were used to inoculate a 250 mL Erlenmeyer flask containing 50 mL of sterile 10-fold diluted LB medium ($LB_{1/10}$). After an overnight culture (30°C, 160 rpm), the bacterial suspension was centrifuged (3,000 g for 4 min) and resuspended in fresh $LB_{1/10}$ medium to reach a final average concentration of $2.3 \times 10^8 \pm 3 \times 10^7$ cells/mL, among which $1 \times 10^6 \pm 2 \times 10^5$ cells/mL were membrane-damaged as measured by flow cytometry (Klopffer *et al.*, 2024).

### Microfluidic system and experiments

The microfluidic system (Fig. 1a) is composed of a high aspect ratio rectangular glass capillary (Vitrocom, height $h$ = 0.1 mm, width $w$ = 1 mm, length $L$ = 100 mm) connected to a syringe pump (NEMESYS Low pressure module). For each assay, a new entire microfluidic device (capillary and tubing) was used and cleaned with the following protocol: SDS (0.1 % - 30 min - 1 mL/min), ultrapure water (30 min at 1 mL/min), HCL (0.1 M - 30 min at 1 mL/min) and ultrapure water (30 min - 1 mL/min). The bacterial suspension fed the system at flow rates $Q$ ranging from 0 to 2 µL/min that led to a good approximation of planar Poiseuille flow far from lateral edges in the $x$-direction (Fig. 1b and Fig. S2): $V_x^{flow}(z) = \frac{6Q}{wh}\left(\frac{z}{h} - \left(\frac{z}{h}\right)^2\right)$. From this velocity profile the wall shear rate is derived: $\dot{\gamma}_w = \frac{6Q}{wh^2}$ (Bruus, 2007). Each experiment lasted less than 2 hours with no bacterial motility variations (Fig. S3).

## Digital holographic microscopy setup and holograms processing for bacteria detection and tracking

The microfluidic system was mounted under an inverted microscope (ZEISS Axio Observer; 40X, 0.9 NA objective) (Fig. 1a). The Gabor in-line holographic setup was composed of a single-mode fiber laser diode (λ = 660 nm, 50 mW, Thorlabs LP660-SF50) fitted with a fiber collimator for plane wave generation (Thorlabs PAF2P-A10A) and mounted above the microscope to illuminate the bacterial suspension within the central part of the capillary (5 cm away from injection point and centered laterally). A digital camera (Dantec Dynamics, pixel size of 7 $\mu m$, resolution 1024 × 1024 pix² matrix) is used to record time-resolved hologram sequences of 3,000 holograms. The frame rate was varied as a function of flow rates in a range [50 - 300] Hz in order to limit maximal bacterial displacements to the bacterial size (2 µm). The field of view was 180 x 180 µm² laterally extended and 100 µm depth.

We recall briefly the hologram formation and analysis with detailed information found in (Becker *et al.*, 2024). The reference electromagnetic wave $E_R$ emitted from the laser is diffracted by the bacteria suspension. The diffracted wave, called the object wave $E_O$, interferes with the reference wave at the sensor plane to form the hologram $H_0(x, y, z = 0, t) = |E_R + E_O|^2$. Before volume reconstruction, the captured holograms are first processed using background division to minimize stationary noise. The background image is computed by averaging the intensity values across the hologram series (3,000 holograms). The cleaned holograms are backpropagated along the optical axis (z – axis) by discrete steps of 0.5 µm over 200 planes to obtain a reconstructed volume of the same size than the field of view (180 x 180 x 100 µm³). The angular spectrum theory (plane waves decomposition) is used as backpropagator kernel which is suited for small propagation distance (Becker *et al.*, 2024). The bacteria coordinates are determined by applying successively for each propagated plane: a Tenengrad focus function to enhance the contrast between bacteria and fluid (Fonseca *et al.*, 2016) and a threshold criterion for image binarization. We then compute a three-dimensional connected component labelling (CCL3D) over the entire volume that group voxels of interest. Finally, the centroid coordinates of each group of voxels are determined and associated to bacteria coordinates. The last step was to build the 3D Lagrangian trajectories from the bacteria positions and this was done with the Trackpy module. All trajectories with less than 10 points are removed from analysis.

Our source code, programmed in Python/Cupy, is in open access (Becker *et al.*, 2024). With our simple informatic equipment (laptop: Core i5, 16GB of Ram, Quadro GTX 3000 GPU), the computation time needed to localize bacteria on a single hologram (backpropagation of 200 planes of 32 floats with a resolution of 1024 x 1024 pix², binarization and CCL3D) was less than 1 second. Then post-processing a time sequence of 3,000 holograms takes around 45 minutes. This is from our knowledge the fastest holograms time computation with a standard laptop and GPU acceleration that allows easy laboratory use.

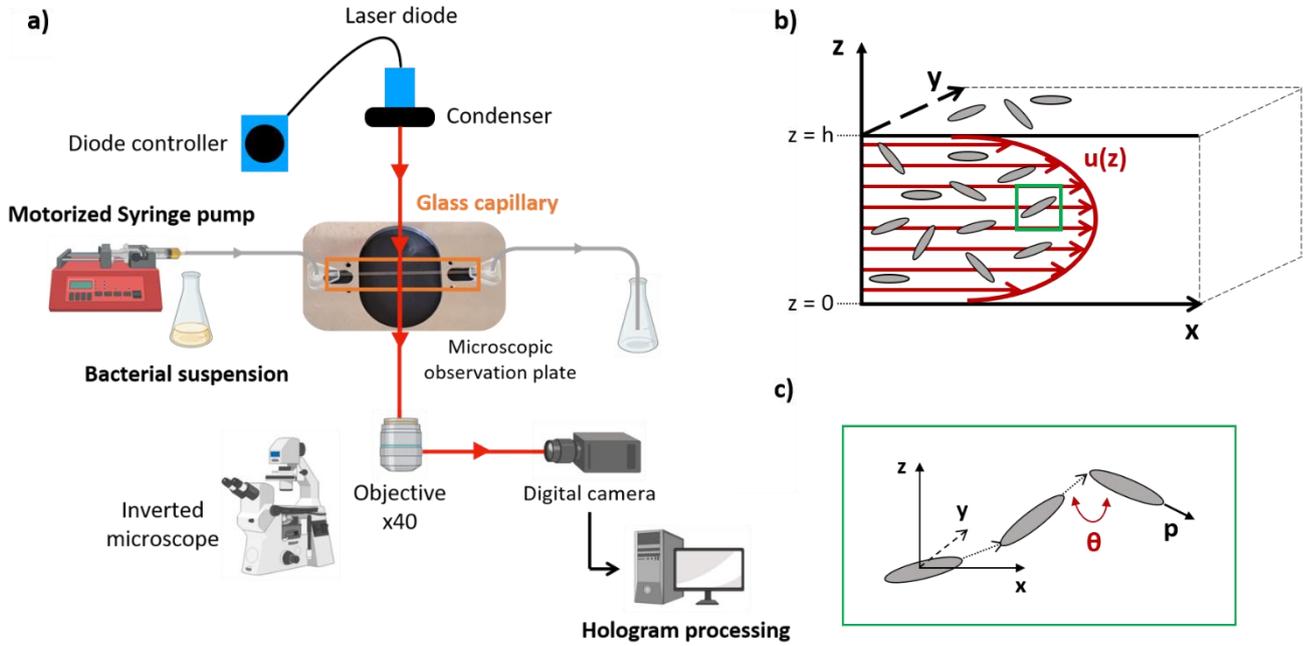

**Fig.1. (a)** Schematic representation of the DHM setup. **(b)** Diagram of the confined capillary system, exhibiting Poiseuille flow. Coordinate axes: *x*-flow direction, *y*- horizontal transverse (vorticity direction) and *z*-vertical confinement direction (shear direction). **(c)** Bacterium unit vector $\vec{p}(t) = \frac{\vec{v}}{\|\vec{v}\|}$ indicating the bacteria orientation and $\theta(t)$ the turning angle.

## Results

### *S. oneidensis* classification dynamics and free-swimming characterization

Our holographic microscopy setup allows for the high-precision three-dimensional reconstruction and analysis of bacterial trajectories, enabling the tracking of 2,000 to 4,000 bacteria per experiment. This method provides valuable insights into the mechanisms underlying bacterial transport properties in inhomogeneous shear flow. Fig. 2 (a-c) illustrates representative 3D Lagrangian trajectories of *S. oneidensis* WT under increasing flow conditions.

The flow strength is characterized by the flow Péclet number: $Pe_f = \frac{\dot{\gamma}_w}{D_r}$ where $D_r$ is the rotational diffusion coefficient that describe the ability of bacteria to keep their orientational swimming direction (see below). This dimensionless number serve as a control parameter and compares the time scale for a bacterium to lose its swimming direction due to active rotational diffusion ($1/D_r$) over the time scale it takes to be aligned by the shear flow gradient ($1/\dot{\gamma}_w$). For $Pe_f \gg 1$ bacteria are aligned in the flow direction and lose the ability to cross streamlines, while for $Pe_f \ll 1$ bacteria swimming dynamic is slightly disturbed by the flow and migration through the channel occurs.

A diverse range of bacterial dynamics is observed through trajectory shapes, displacement angles and explored volumes that calls for a classification. Based on simple criteria (see partie thèse) we determined three categories (Fig.2 (d-g)) and their evolution with $Pe_f$. The first category (C1) 'diffused bacteria', which are either wall-adhered or exhibit Brownian motion, (C2) 'convected bacteria' characterized by trajectories following the streamlines and (C3) 'motile bacteria' that cross the streamlines.

As the flow strength increases, the C1 category drops rapidly (Fig. 2g). While in static or low flow rates conditions C1 trajectories are in the entire volume, moderate or high shear localized this category near walls. Indeed, non-motile bacteria far for the walls are carried by the flow and fall in the category C2. The fraction of motile bacteria (C3) also evolves with the imposed flow rate. While the fraction is close to 80 % in static condition, it drops to 20% of the entire population when a shear is applied and 10% for the larger flow rate. Their movement can be attributed solely to the presence of a synthesized flagellum, as such trajectories are absent in the *S. oneidensis* ΔFla mutant strain (Fig. S4). Finally, at large shear, most of bacteria are convected.

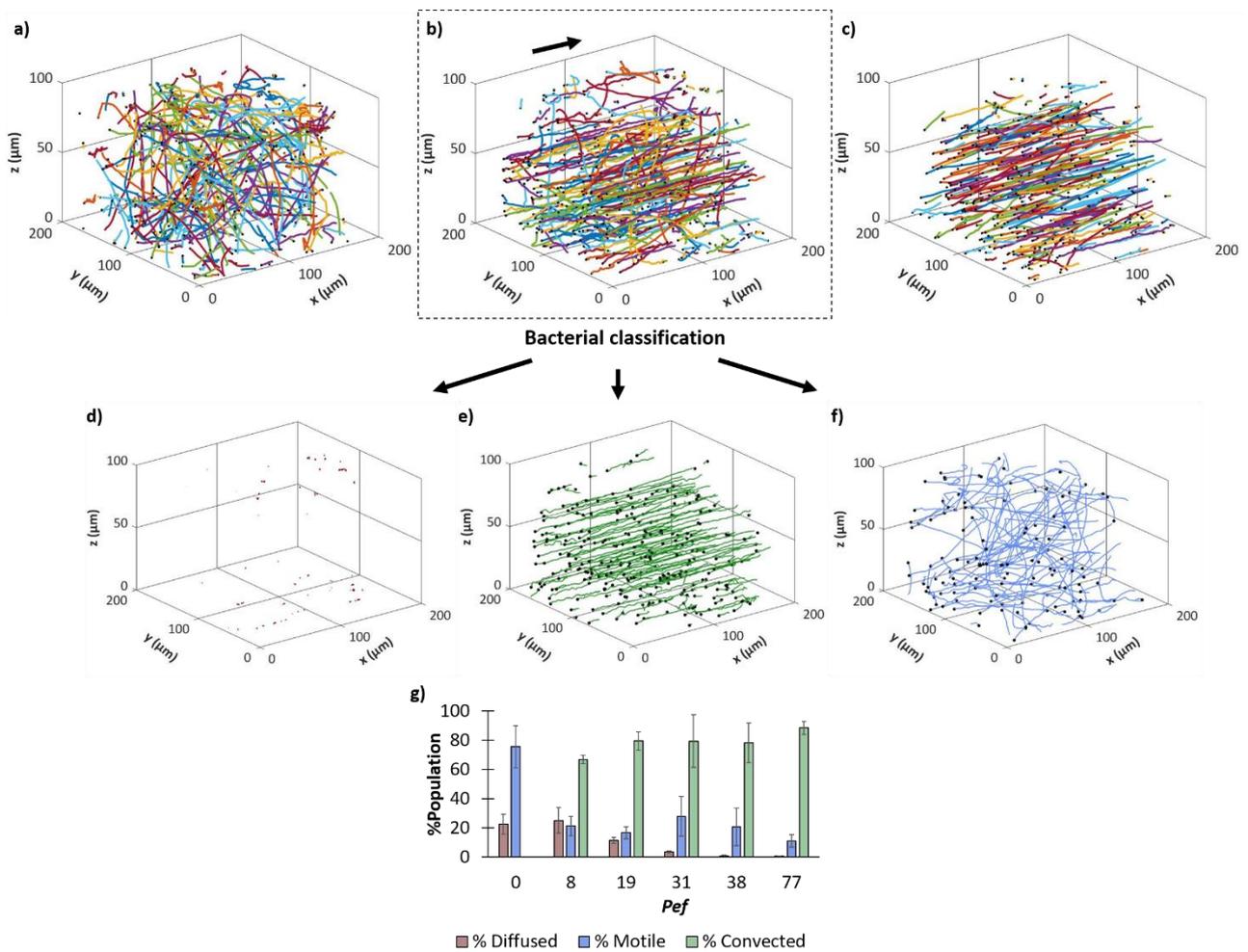

**Fig.2. (a)**, **(b)** and **(c)** are representative examples of three-dimensional trajectories (about 500) of wild type *Shewanella oneidensis* obtained for different values of flow conditions **(a)** $\dot{\gamma}_w = 0$ s$^{-1}$, $\langle V_x^{flow} \rangle = 0$ µm/s, **(b)** $\dot{\gamma}_w = 2$ s$^{-1}$, $\langle V_x^{flow} \rangle = 33$ µm/s, **(c)** $\dot{\gamma}_w = 20$ s$^{-1}$, $\langle V_x^{flow} \rangle = 330$ µm/s) within a 180 × 180 × 100 µm³ volume. The capillary walls are located at $z = 0$ (bottom) and 100 µm (top). Flow direction is along the positive x-axis, as indicated by the black arrow in figure **(b)**. Figures **(d)**, **(e)** and **(f)** are representative examples of bacterial trajectories obtained after classification of those obtained at $\dot{\gamma}_w = 2$ s$^{-1}$: **(d)** C1 'diffused bacteria', **(e)** C2 'convected bacteria' **(f)** C3 'motile bacteria'. **(g)** Mean ± standard deviation of proportion of populations identified by transport mechanism as a function of flow Peclet number.

To better understand the transport properties of *S. oneidensis*, we focus on the motile population and characterized, without external flow, their swimming properties in terms of reorientational angle, velocities distributions and orientational correlation (partie thèse for detailed calculations). Fig. 3a displays a selection of 10 out of 12,241 motile *S. oneidensis* trajectories. These trajectories illustrate distinct behaviors, although in varying proportions (Fig. S5). The predominant behavior, accounting for 70% of the population of motile bacteria, consists of long and slightly curved runs characterized by low reorientation angles (bacteria 1 to 4). Also observed to a lesser extent, bacteria that performed large reorientation ($\theta > 50°$) which we associate to tumbles (bacteria 5 to 7). At last, few percent of the bacteria population exhibit low swimming speed (bacteria 8) (Molaei *et al.*, 2014) and near surface behavior (bacteria 9 and 10) that was previously observed in *E. coli* (Kaya and Koser, 2012; Lauga *et al.*, 2006; Li *et al.*, 2008a). These observations align with the only study on *S. oneidensis* MR1 performed by Stricker *et al.* (2020) which demonstrated that this bacterium can exhibits multi-modal swimming behaviors like run-reverse-flick and run-and-tumble movements.

The bacterial turning angle (Fig. 3b) and instantaneous swimming velocities (Fig. 3c) supports our previous observations. The distribution of turning angle reveals the predominance of small angle reorientations (0 - 20°) consistent with the *S. oneidensis* motility pattern of long runs with few directional changes, yielding a mean turning angle of $\langle \theta \rangle \approx 35°$, with a substantial standard deviation of 45°, reflecting the presence of angles larger than 50°. Associated to these smooth trajectories, the intrinsic bacterial swimming velocities $v_0$ appears bimodal with a high velocity peak centered at 50 µm/s associated to bacteria swimming in the bulk and a low velocity peak around 5 µm/s for those interacting with a wall (Fig. S6). To go further on the swimming properties, we analyze the three components of the bacterial velocity (Fig. 3d). While the velocity components in the plane perpendicular to the confinement direction are similar, the z-component distribution is narrowed and less homogenously distributed. We assume that this effect is related to confinement effect and to the large swimming persistence of *S. oneidensis*. Indeed, the calculation of the directional correlation function $C(\Delta t) = \langle \vec{p}(t) \cdot \vec{p}(t + \Delta t) \rangle = e^{-2D_r t}$, where $\vec{p} = \frac{\vec{V}}{\|\vec{V}\|}$ is the orientation of the bacteria, leads to rotational diffusion coefficient $D_r \approx 0.25$ rad²/s and a correlation time $t_c = \frac{1}{2D_r} \approx 2$ s (Fig. 3e). This demonstrate that during that time a motile bacterium performs a smooth trajectory with a persistent length $l_p = v_0 t_c \approx 100$ µm which is of the same order of magnitude as the confinement ($h = 100$ µm). This long correlation time (or low rotational diffusion coefficient) is also found in other monoflagellate bacteria like *Rhodobacter sphaeroides* which has a similar coefficient (0.255 rad²/s) (Rosser *et al.*, 2014). Consequently, most of the bacteria encounter a wall during their runs before performing a tumble as it is highlighting in the inset of Fig. 3b where large turning angles occur at the wall rather than in the bulk. This behavior is quite different form the typical peritrichous run & tumble *E. coli* vastly studied in the literature (Berg and Brown, 1972; Lemelle *et al.*, 2020; Saragosti *et al.*, 2012). For this bacteria, reported values $D_r \approx 1 - 2$ rad²/s and $v_0 \approx 20 - 30$ µm/s lead to a persistence length of 10 µm allowing for the bacteria to mostly tumble in the bulk. This distinction suggests different transport dynamics and highlights the need for caution when comparing numerical models of bacterial transport in confined environments, as these models predominantly rely on *E. coli* data (Bhattacharjee and Datta, 2019). Therefore, incorporating observations from our strain into models can improve the accuracy of transport predictions by accounting for motility differences.

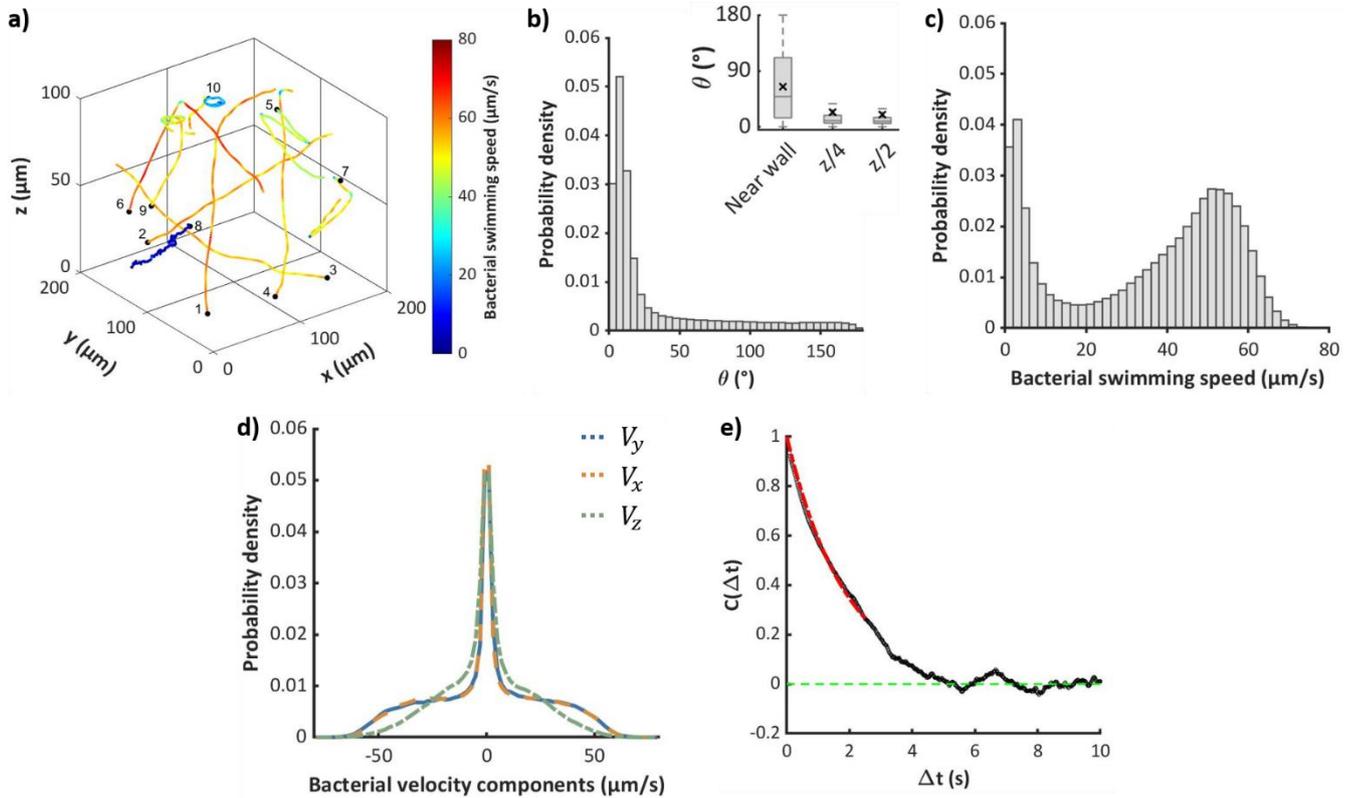

**Fig. 3.** Swimming dynamics properties of the motile population part of *Shewanella oneidensis WT* without flow. **(a)** ten representative examples of the 12,241 motile bacteria color encoded by their instantaneous velocities. The capillary walls are located at z = 0 (bottom) and z = 100 μm (top). **(b), (c),** and **(d)** show, respectively, the probability density function of: the turning angle **θ** (inset: distribution of the turning angle as a function of the position inside the capillary), the instantaneous bacterial swimming speed and the bacterial velocity component in each direction. **(e)** Mean directional correlation function fitted exponentially (red curve) to obtain persistence parameters ($t_c$ = 1.9 s; $D_r$ = 0.26 rad²/s).

### Shear flow impact the transport properties of *S. oneidensis*

- **Flow effects on bacterial 3D trajectories and velocities**

Imposing a flow on a bacterial suspension strongly affect their transport properties. Examples trajectories of *S. oneidensis* under various $Pe_f$ conditions are depicted on Fig. 4 (a-c) (Fig. S7 for more trajectories). For the lowest flow strength ($Pe_f = 8$), while the fraction of motile bacteria drops to 20% (Fig. 2g) their behavior is similar to the static case and keep the ability to cross streamlines with persistent runs, tumbles and surface swimming. As the $Pe_f$ increases, bacteria trajectories tend progressively to align with flow streamlines reducing the ability to explore their environment. Our observations are supported by the narrowing of the turning angle distribution and the decrease by a factor of 2 of the mean (Fig. 4e).

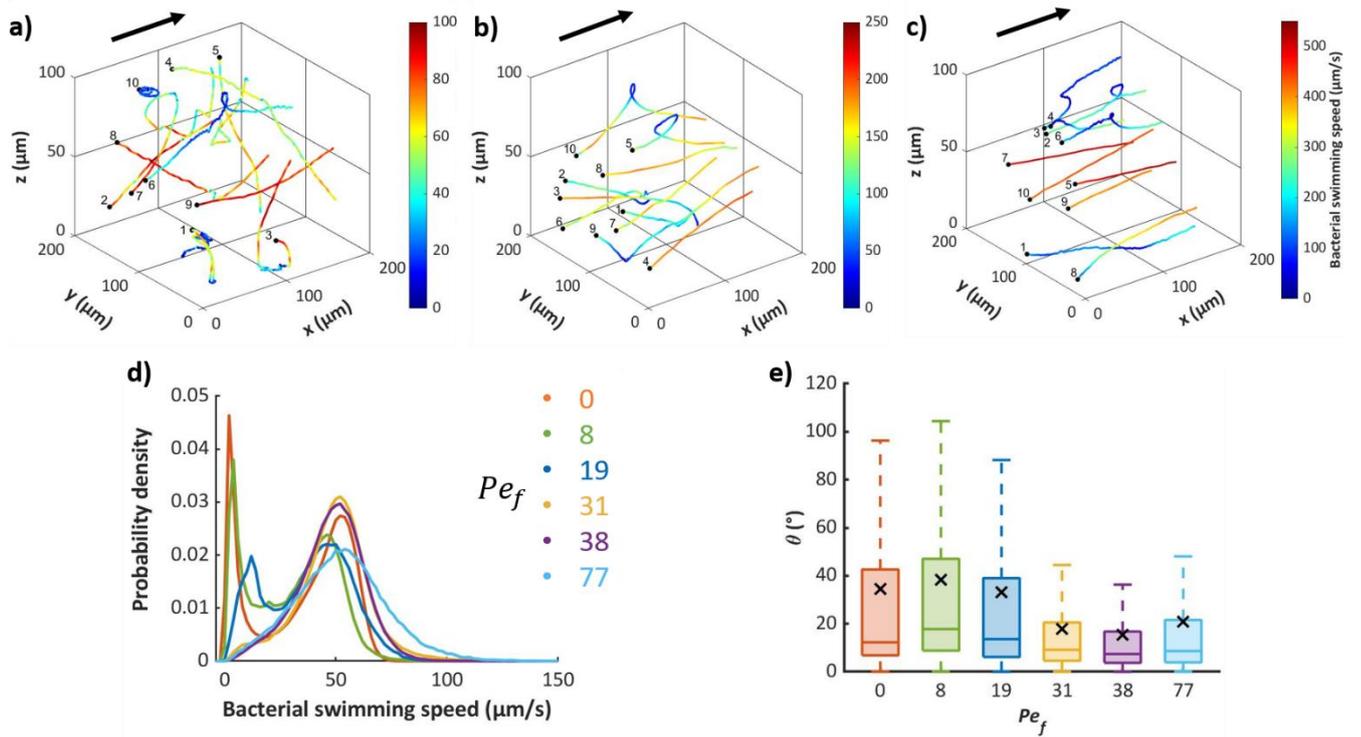

**Fig. 4. (a)**, **(b)** and **(c)** show 10 representative examples of three-dimensional trajectories of motile wild-type *Shewanella oneidensis* obtained under different flow Peclet number conditions (a = 8, b = 31, c = 77). The capillary walls are located at z = 0 (bottom) and 100 µm (top). Flow direction is along the positive x-axis, as indicated by the black arrow. **(d)** Distribution of motile *S. oneidensis* intrinsic bacterial swimming speed for various Peclet number. **(e)** Boxplot of the turning angle θ as a function of the flow Peclet number applied. The number of motile bacteria analyzed was 28,114 (with 12,241 for $Pe_f$ = 0; 1,816 for $Pe_f$ = 8; 1,991 for $Pe_f$ = 19; 5,346 for $Pe_f$ = 31; 4,192 for $Pe_f$ = 38 and 2,528 for $Pe_f$ = 77).

Interestingly, this drastic change in bacterial trajectories does not appear to be accompanied by any significant variation in bacteria's intrinsic swimming speed once the flow contribution is removed (Fig. 4d). In line with Chattopadhyay *et al.* (2006), this result suggests that, despite the increased drag forces, bacteria maintain their energy for active swimming, preserving their own propulsion capabilities. Indeed, Fig. 4d shows that swimming speed distributions remain similar across shear flows, with only minor deviations. In particular, as the flow increases, the low-velocity peak - primarily due to near-wall swimming- diminishes and disappears once the Peclet number exceeds 19. This reduction in low-velocity events is linked to the decline, under increasing wall shear rates, of complex near-wall trajectories such as upstream swimming and surface rheotaxis (Berke *et al.*, 2008; Drescher *et al.*, 2011; Kaya and Koser, 2012; Lauga, 2016; Lauga *et al.*, 2006; Li *et al.*, 2008b; Mathijssen *et al.*, 2019). As a result, the original bimodal speed distribution is transformed into a unimodal distribution centered on highest swimming velocities.

- **Flow impact bacterial spatial orientation and distribution through the channel**

We now turn on a more quantitative description of the three-dimensional spatial orientation of *S. oneidensis* under flow. Fig. 5 (a-c) (and Fig. S8) shows, for various Peclet number, the distribution of each component of the orientation vector $\vec{p} = (m_x, m_y, m_z)$ respectively in the flow direction - x, the vorticity direction - y and the wall normal direction - z. We notice that for all the results presented below, we did not subtract the flow component leading to orientation vectors measured in the laboratory reference frame. We first notice that the component $m_y$, $m_z$ stay isotropic and narrow as $Pe_f$ increases. Nevertheless, compared to the Δfla strain, the distributions are broader indicating rheotactic effects that are more pronounced for $Pe_f > 19$ as we will show later (Fig. S9, Fig. 6). The streamwise component (Fig. 5a) reveals, at moderate Peclet number, that upstream motion occurs for a large part of the bacteria while the Δfla strain is totally convected ($m_x = 1$) (see Fig. S10 for the $V_x$ distribution as function of the Péclet number). The spatial analysis (Fig. 5d) shows that while the average $<m_x>$ increases with flow throughout the capillary, this increase is more pronounced at the center than near the walls. Notably, a boundary layer around 10 µm adjacent to the walls maintains $<m_x>$ below 1, even at high shear, indicating that bacteria near walls experience hydrodynamic interactions like surface rheotaxis (Figueroa-Morales *et al.*, 2020; Mathijssen *et al.*, 2019).

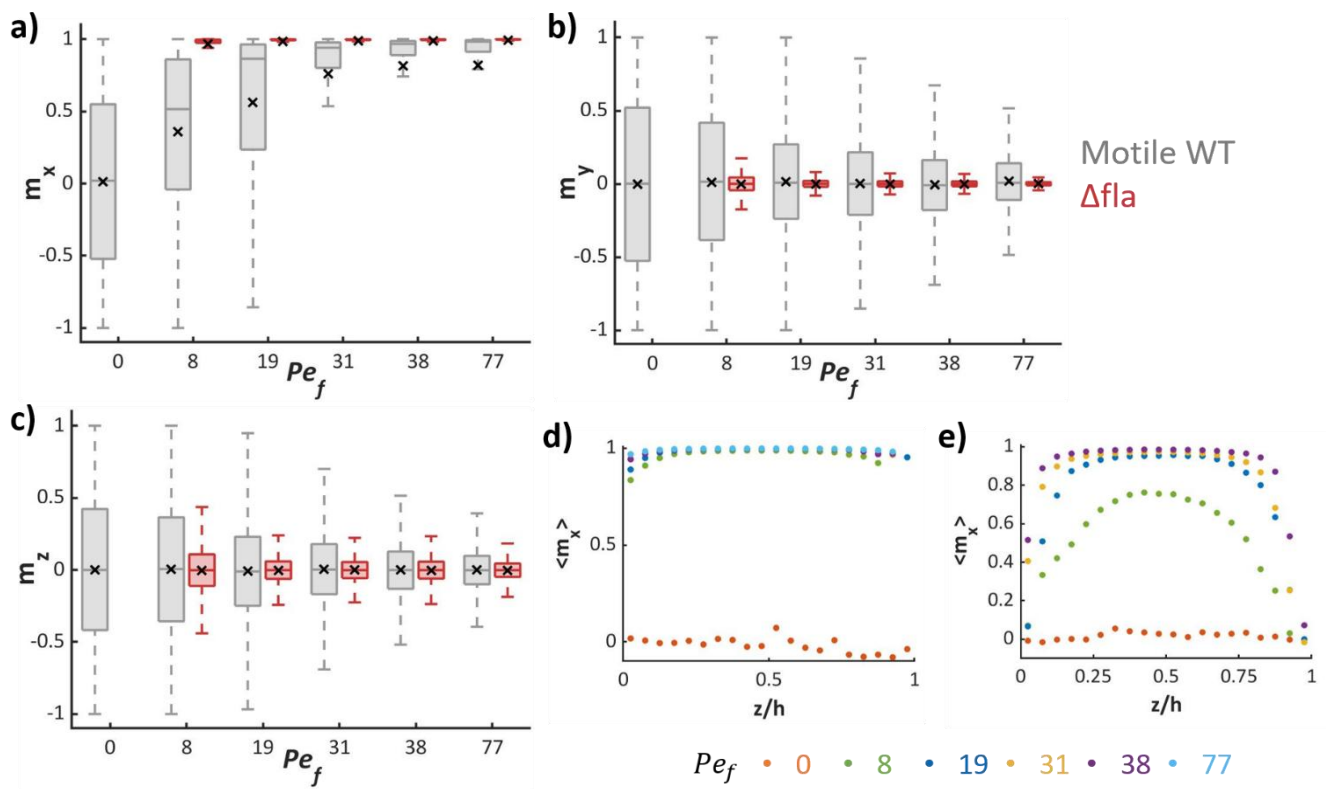

**Fig. 5.** Boxplots of bacterial orientation **(a)** x-direction, $m_x$ **(b)** y-direction, $m_y$ and **(c)** z-direction, $m_z$ for motile *S. oneidensis* WT (grey) and Δfla mutant (red) strains under applied flow Peclet number. **(d)** and **(e)** Plots showing the average x orientation ($m_x$) of the ΔFla mutant **(d)** and the motile population of *S. oneidensis* WT **(e)** under applied flow Peclet number as a function of their normalized position within the capillary. A total of 28,114 bacteria were analyzed for the motile WT population, and 84,103 for the Δfla mutant.

This complex orientational behavior induces a coupling with the bacterial concentration profile inside the capillary. As seen in Fig. 6a (Fig. S11a for Δfla strain), the concentration profiles are sensitive to the flow strength. In static condition ($Pe_f = 0$), the concentration is homogeneous in the bulk but increases drastically close to the wall by a factor of 3. This accumulation effect is well known for pusher bacteria. Indeed, some studies (Berke *et al.*, 2008; Clement et al., 2016; Drescher et al., 2011)) argue that this accumulation comes from hydrodynamic interactions between the far velocity field induced by the self-propulsion of the bacteria and the walls. In that case, bacteria tend to align with the surface resulting in a zero orientation toward the wall. Another explanation proposed by some authors (Ezhilan and Saintillan, 2015) asserts that near a wall, the flux balance between self-propulsion $cv_0$ and diffusion $\frac{D_t c}{\delta}$ lead to orientational sorting that drives the accumulation of bacteria and a net polarization toward the surface. This balance leads to an estimation of the accumulation thickness $\delta^{th} = \frac{D_t}{v_0}$. Taking $D_t \approx \frac{v_0^2 t_c}{3}$ (Lovely and Dahlquist, 1975) and $v_0 \approx 40$ µm/s and $t_c = 2$ s from our results, we find $\delta^{th} \approx 25$ µm. From the concentration profile we estimate the accumulation layer ($c\left(\frac{z}{h} = \frac{\delta}{h}\right) = 1$) to $\delta \approx 5$ - 10 µm corresponding to few cell sizes and is significantly smaller than $\delta^{th}$. Also, we show on Fig. S9d that the average orientation toward the wall is homogeneous in all the capillary and equal to zero.

Contrary to what is mentioned in Ezhilan and Saintillan (2015) and to our knowledge, there are no experimental findings reporting under the conditions for which the model was developed (diluted suspension) that, close to a surface, a clear orientation toward the wall exist. We thus agree with the first explanation as our results suggest this trend. We can however moderate our conclusion because of the strong persistent swimming properties of *S. oneidensis*. Indeed, previous works mainly focus on *E. coli* (Berke *et al.*, 2008; Li *et al.*, 2011; Li and Ardekani, 2014; Li and Tang, 2009; Molaei *et al.*, 2014) which is a typical case of run & tumble behaviour for which the effective diffusion coefficient is smaller and could be the reason of discrepancies.

When applying a flow, the concentration profile slightly evolves and tends to flatten with two particular effects. We first observe that the wall concentration decreases with the Péclet number. Secondly, the concentration in the near-wall region increases and growth compared to the static case. A rough estimation, for $Pe_f > 19$, gives a thickness that is increased up to 20 µm. This effect, called shear trapping, was theoretically predicted (Ezhilan and Saintillan, 2015; Ganesh *et al.*, 2023) but rarely experimentally observed (Rusconi *et al.*, 2014). Thanks to holography, we are able to characterise this effect more precisely by measuring the average bacterial velocity component normal to the wall $<V_z>$ and its evolution between surfaces (Fig. 6b inset). Surprisingly, we observe for each flow strength two peaks that are symmetrically located on either side of the centreline with opposite magnitudes. As the Péclet number increases, the peaks get closer to the center of the channel and increase in magnitude varying between 2 and 10 µm/s. This result indicates a net migration of bacteria away from the center that induces a growth of the region where shear trapping occurs. In order to rationalize this evolution, we plot on Fig. 6b (Fig. 11b for Δfla strain) $<V_z>$ as a function of the local Péclet number $Pe(z) = \frac{\dot{\gamma}(z)}{D_r}$. With the experimental precision, the peaks aligned at the same value of $Pe(z) \approx 7$ indicating that the maximal migration velocity only depends on the shear rate. In the center, between the peaks, $<V_z>$ goes to zero due to the predominance of the self-propulsion compared to shear alignment that vanishes as bacteria approach the centreline.

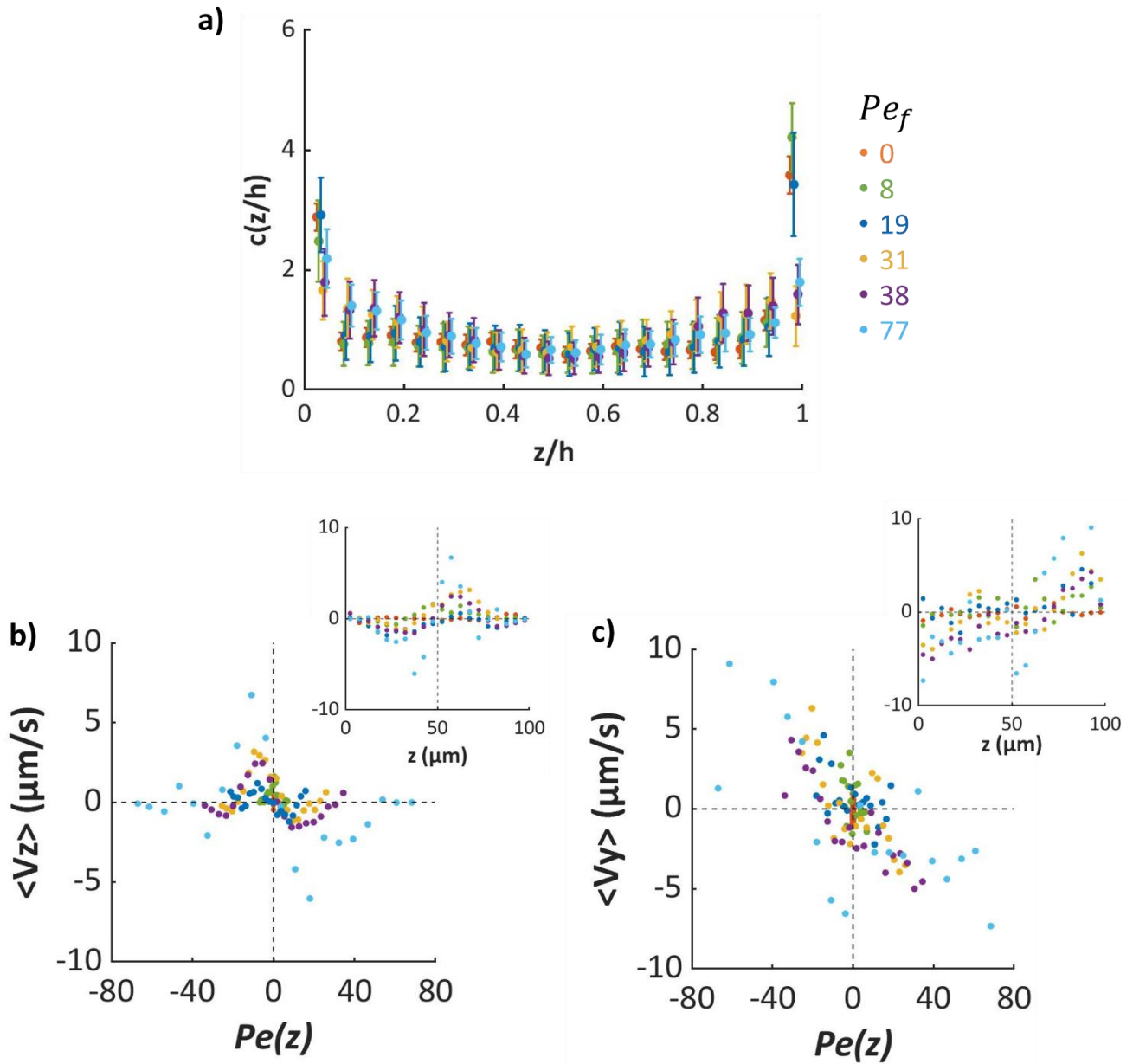

**Fig. 6. (a)** Vertical concentration profiles of *S. oneidensis* WT from the bottom wall as a function of the normalized distance z across the capillary height. Curves are color-coded by the corresponding flow Peclet number. **(b-c)** Mean bacterial velocities: $V_z$ (µm/s) in panel (b) and $V_y$ (µm/s) in panel (c) as a function of the applied local Peclet number ($Pe_f$) for motile *S. oneidensis* population. Insets show the same mean velocity measurements but as a function of the z position within the capillary. A total of 28,114 bacteria were analyzed.

This led to isotropic bacteria orientation. Beyond those peaks, $<V_z>$ rapidly drops to zero indicating in that case that the bacteria are aligned by the flow gradient and stay captured with a reduced possibility to cross streamlines. In the work of Ezhilan and Saintillan (2015), these trends are theoretically demonstrated but they report a non-linear dependence of the position of the peaks with the flow Péclet number $Pe_f$. It is interesting to note that, while close to the wall our experimental results differ from their work, their model captures our experimental trends in the bulk region of the channel.

To get a complete view of the bacterial suspension dynamic under flow we also characterize the average transverse velocity $<V_y>$ of S. oneidensis (Fig. 6c and Fig. 11c for Δfla strain). This monotrichous bacterium shows bulk rheotaxis that depends only of the local shear rate as it is shown in Fig. 6c. The rheotactic velocity increases with local shear rate (local Pécet number) and account for 20% of the intrinsic swimming speed of the bacteria that is in-line with experiments on peritrichous bacteria (Jing *et al.*, 2020; Marcos *et al.*, 2012). Without quantitative explanation, we notice that the rheotactic velocity $<V_y>$ is quite similar to the velocity normal to the walls $<V_z>$. This could be interpreting as an equilibrium between bacteria migration from the centreline and bacteria entering by bulk rheotactic effect. This assumption will be explored in a future study with the quantification of the three-dimensional bacterial flux.

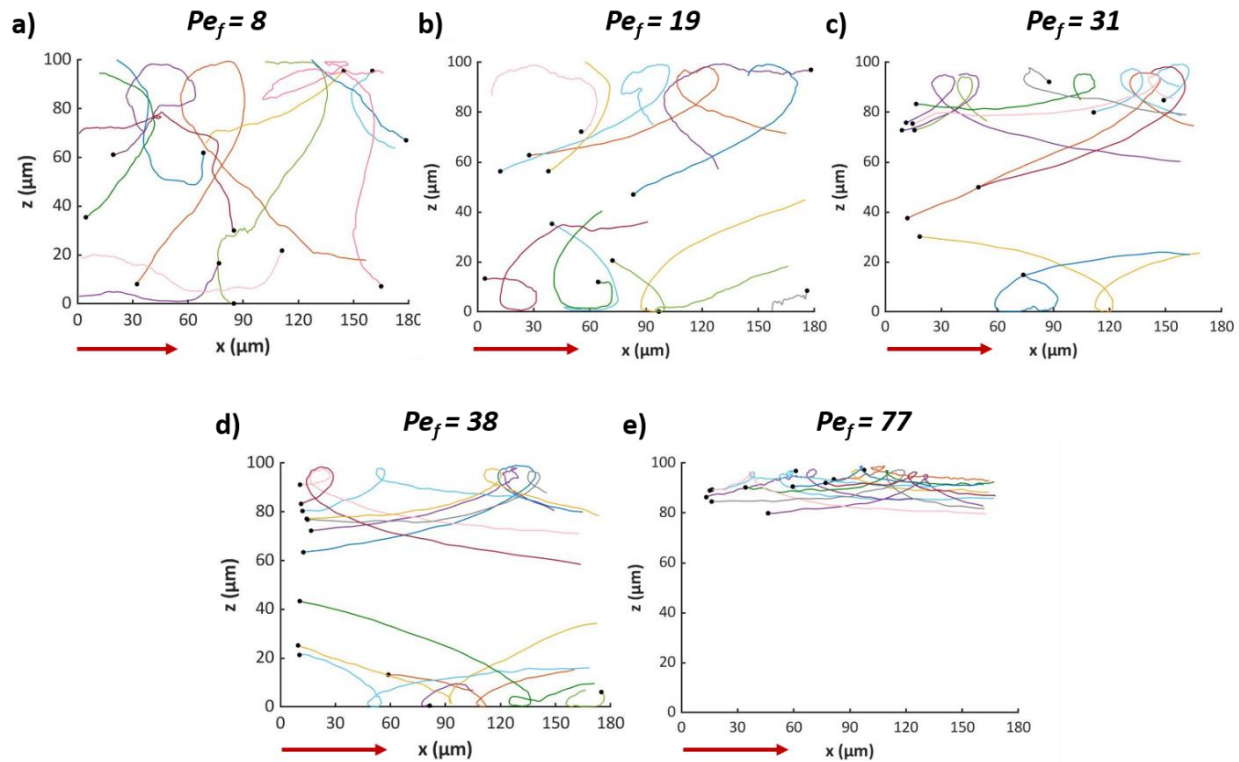

**Fig. 7.** Representative trajectories of motile *S. oneidensis* projected onto the z-x plane, observed under the different Peclet numbers applied. These trajectories illustrate Bretherton-Jeffery orbits. Trajectories are color-coded individually; black circles mark the initial bacterial positions.

Finally, we have tried to understand what could explain our result $<V_z> = 0$ close to a wall and the discrepancy observed with the theoretical model of Ezhilan and Saintillan (2015). A deeper analysis of the Lagrangian trajectories in the neighbour of the surfaces highlight a large proportion of bacteria experiencing looping even at high $Pe_f$ (Fig. 7). These looping, called active Bretherton-Jeffery orbits, was theoretically expected first by Zöttl and Stark (2012) and experimentally reported one time for a non-tumbling *E. coli* (Junot *et al.*, 2019). We do not describe in detail these active orbits but only put forward that along a trajectory, bacteria experience a velocity toward the wall when entering the loop ($V_z > 0$) and a velocity against the wall when leaving the loop ($V_z < 0$) with the same magnitudes. Averaging over thousands bacteria trajectories this results in a cancelation of the net orientation. As the $Pe_f$ increases, the fraction of motile bacteria that experience active orbits decreases (Fig. S12) and $<V_z> = 0$ seems to be obtained by shear alignment in the flow direction.

# Conclusion

Understanding how bacteria evolve in their environment is crucial for predicting dispersion and surface colonization. In this study, we experimentally investigated the behavior of a dilute suspension of *Shewanella oneidensis* MR-1 in a confined Poiseuille flow. We show that this monotrichous species interacts strongly with boundaries due to its highly persistent swimming, and couples to the flow in a manner distinct from run-and-tumble bacteria such as *E. coli*. Our ability to analyze thousands of individual trajectories enabled us to obtain robust measurements of the spatial distribution of *S. oneidensis* in flow. We found that while kinetic models capture migration and shear trapping far from walls, they fail to describe the net bacterial orientation near surfaces. Although individual cells exhibit nonzero scattering angles upon interacting with the wall, this effect vanishes at the population level, particularly as wall shear rate increases. These results support the view that complex hydrodynamic interactions between bacteria and surfaces must be considered. Finally, our digital in-line holographic microscopy setup proved to be a rapid and powerful tool for characterizing suspensions containing both motile and non-motile bacteria, even under high flow conditions. We argue that gaining insight into pioneer surface colonization requires studying bulk-to-surface transfer mechanisms at the population scale-an approach that DHM readily enables.

# Acknowledgements

We would like to express our sincere gratitude to Cécile Jourlin-Castelli (BIP, Aix-Marseille Université) for kindly providing the *S. oneidensis* Δfla strain used in this study. This work was supported by the French National Research Agency through the 'BIOCIDES' project (ANR-21-CE50-0027). N. L. acknowledged the french France 2030 program "Lorraine Initiative of Excellence", reference ANR-15-IDEX-04-LUE for supported partly this work.

# Appendix A. Supplementary data

## Bacterial trajectory analysis

- The bacterial velocity in the laboratory reference frame $\vec{V}(t) = (V_x, V_y, V_z)$ are obtained from the convolution of the spatial coordinates along the trajectory $\vec{x}(t)$ with the first derivative of a Gaussian kernel (**1**). The instantaneous velocity is calculated as $\sqrt{V_x^2 + V_y^2 + V_z^2}$.

- The instantaneous orientation is determined from the velocity vector in the laboratory reference frame $\vec{p}(t) = \frac{\vec{V}(t)}{\|\vec{V}(t)\|}$.

- The mean square displacement used to determine the motility of *S. oneidensis* (Figure S1) is computed as follow:

$$MSD(\Delta t) = <[x(t+\Delta t) - x(t)]^2 + [y(t+\Delta t) - y(t)]^2 + [z(t+\Delta t) - z(t)]^2>$$

- All average quantities presented in the document are ensemble averages over all the bacterial trajectories.

## Bacterial trajectory classifications

- <u>C1 category:</u> Bacteria that are attached to a wall or diffusing in the flow due to Brownian motion only are considered as "diffused" category (see text). We define a simple criterion based on the length of diffusion $L_{Diff} = \sqrt{D_B T}$, where $D_B = 0.7$ µm²/s is the Brownian diffusion coefficient for a passive bacteria (ref) and $T$ is the trajectory recording time. If the distance of the most distant point from the beginning of the trajectory is less than $L_{Diff}$, then the bacteria fall in the "diffused" category.

- <u>C2 category:</u> convected bacteria have straight trajectories without migration across streamlines. We define a metric that compares the curve length $L$ to the end-to-end vector $\Delta$ of the trajectory:

$$L = \sum_{i=1}^{n-1} \sqrt{(x(i+1) - x(i))^2 + (y(i+1) - y(i))^2 + (z(i+1) - z(i))^2}$$

and

$$\Delta = \sqrt{(x(final) - x(initial))^2 + (y(final) - y(initial))^2 + (z(final) - z(initial))^2}$$

If the three following conditions are satisfied: $\frac{L}{\Delta} < 5$ and $|y(final) - y(initial)| < 10$ µm and $|z(final) - z(initial)| < 10$ µm, bacteria are considered as convected by the flow.

- <u>C3 category:</u> Bacteria that enter in the "motile" category are those that were excluded from C1 and C2 categories.

## References

1. Mordant, N., Crawford, A. M., & Bodenschatz, E. (2004). Experimental Lagrangian acceleration probability density function measurement. *Physica D: Nonlinear Phenomena*, *193*(1-4), 245-251.

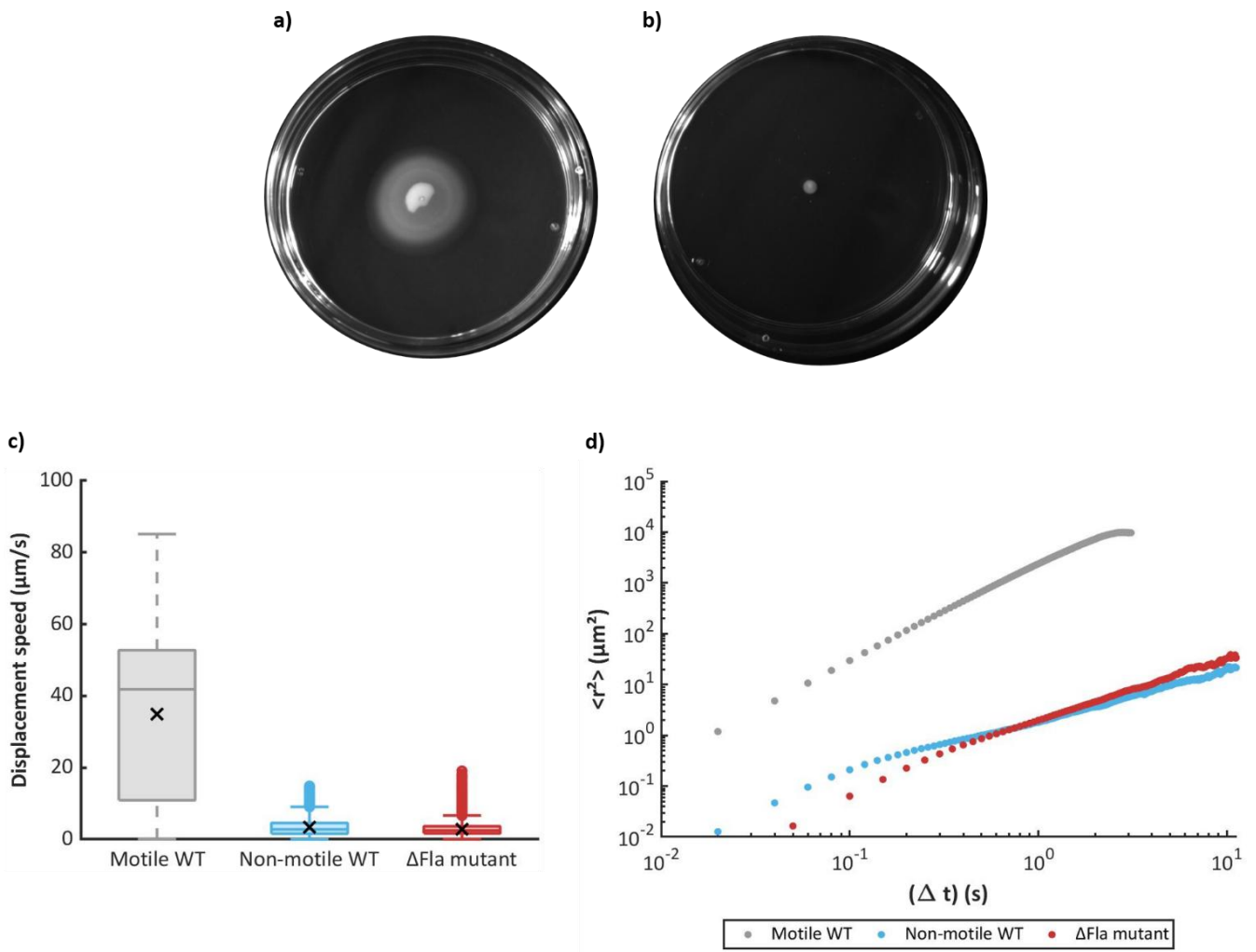

**Fig. S1. (a-b)** Migration behavior of *Shewanella oneidensis* MR-1 WT **(a)** and MR-1 Δfla deletion mutant **(b)**. Motility assays were performed in LB plates containing 0.2% agar at a bacterial concentration close to $2\times10^8$ cellules/mL. After inoculation with 2 µL of the bacterial suspension, the plates were incubated at room temperature (22 ± 0.5°C) for 4 days. The pictures below were taken after 42 hours. Three independent trials were carried out before using these strains in our study.

**(c)** Boxplots representing the bacterial speed distributions of the different bacterial populations identified in *S. oneidensis*: motile and no-motile wild type strains, and the Δfla mutant in static conditions. The central line represents the median; boxes indicate the interquartile range, and the black cross represents the mean value. **(d)** Mean squared displacement (MSD) curves for these bacterial populations. The number of bacteria analyzed are 12,241 for the motile wild-type, 8,110 for the non-motile wild-type, and 5,986 for the Δfla mutant strain.

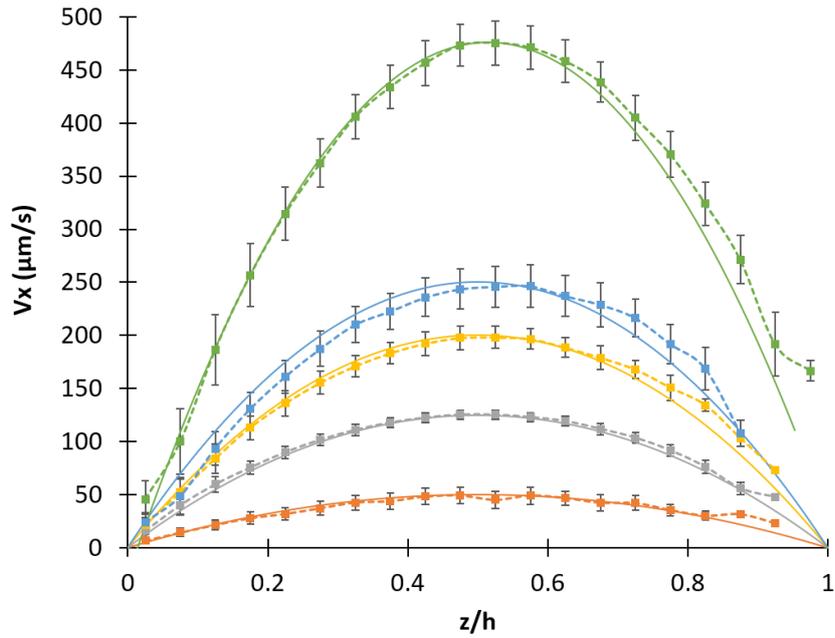

**Fig. S2.** Velocity profiles $V_x^{flow}(\frac{z}{h})$ obtained by holographic tracking of passive tracer particles (2 µm latex beads) in the experimental system at various applied shear rates: 2 s⁻¹ (orange), 5 s⁻¹ (grey), 8 s⁻¹ (yellow), 10 s⁻¹ (blue), and 20 s⁻¹ (green). The flow is directed along the x-axis. The vertical position z represents the particle position within the capillary and is normalized by the total capillary height (100 µm). Experimental data are shown as symbols with dotted lines, while solid lines represent the theoretical Poiseuille flow profile.

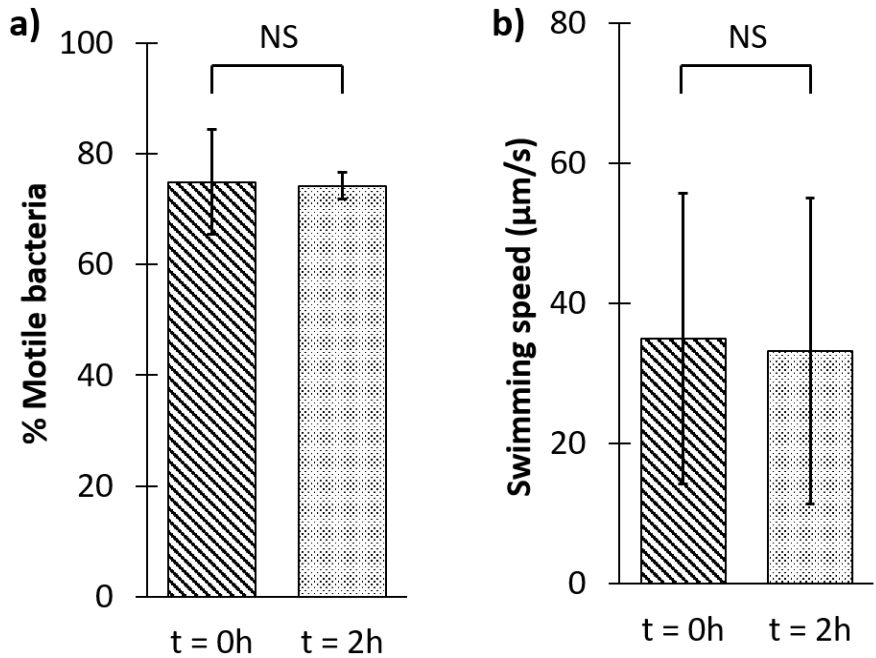

**Fig. S3.** Mean proportion of *S. oneidensis* MR-1 WT motile bacteria (a) and swimming speed (b) at the beginning (t = 0 h) and the end (t = 2 h) of the experiment. Error bars indicate standard deviation from four independent experiments. The total number of bacteria analyzed was $n_{total}$ = 15,051 (with 7,051 at t = 0 h and 8,000 at t = 2 h). (NS = not statistically significant).

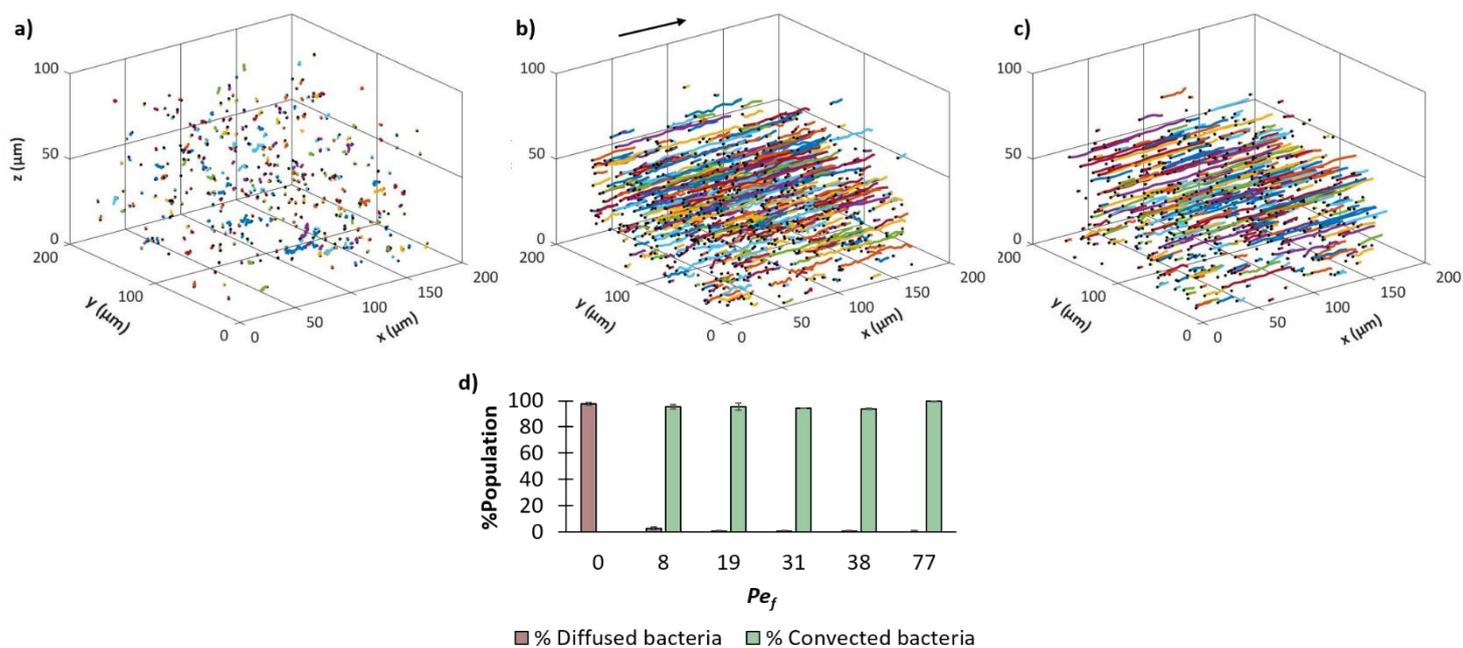

**Fig. S4.** Graph **(a), (b)** and **(c)** are representative examples of three-dimensional trajectories (about 500) of *Shewanella oneidensis* MR-1 Δfla mutant obtained for different flow Peclet number (a = 0, b = 8, c = 38) within a 180 × 180 × 100 µm³ volume. The capillary walls are located at z = 0 (bottom) and 100 µm (top). Flow direction is along the positive x-axis, as indicated by the black arrow in figure b. **(d)** Mean ± standard deviation of proportions populations identified by transport mechanism as a function of applied flow Peclet number. The number of total bacteria analyzed was 88,875 (with 14,704 for $Pe_f$ = 0; 14,727 for $Pe_f$ = 8; 20,703 for $Pe_f$ = 19; 8,703 for $Pe_f$ = 31; 18,389 for $Pe_f$ = 38; 6,877 for $Pe_f$ = 77). Data were obtained from 2 independent experiments per shear rate.

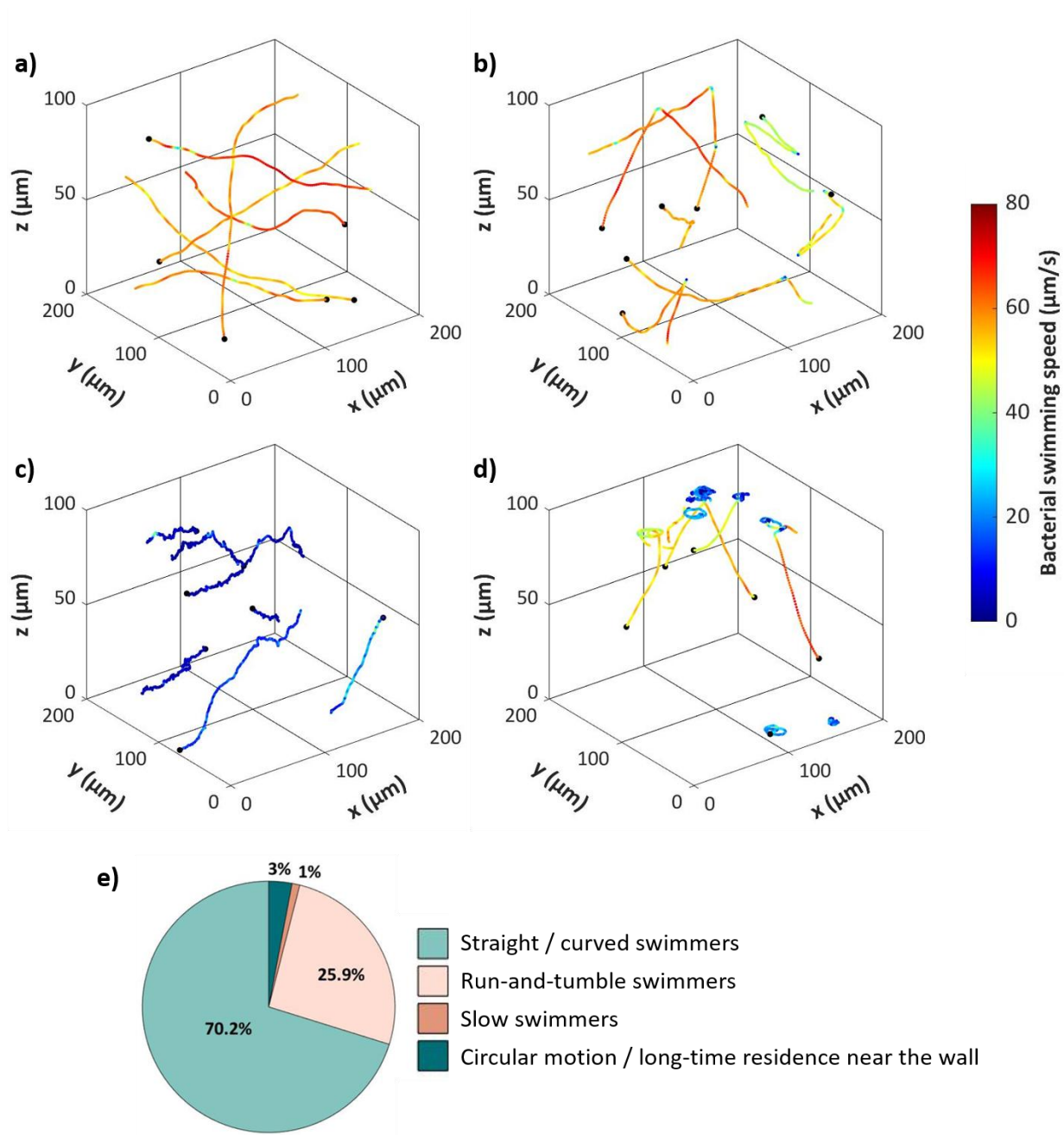

**Fig. S5.** Representative examples of three-dimensional swimming trajectories of the wild-type *S. oneidensis* strain within a 180 × 180 × 100 µm³ observation volume, acquired under static conditions. Trajectories are color-coded according to the instantaneous swimming speed (see colorbar on the right) and the starting point of each trajectory is indicated by a black dot. The capillary walls are located at z = 0 µm (bottom) and z = 100 µm (top). Sample trajectories illustrate distinct swimming behaviors: **(a)** long, straight or gently curved runs; **(b)** run-and-tumble dynamics, characterized by abrupt directional changes (θ > 50°); **(c)** slower, less directed swimming; and **(d)** near-wall circular motion or prolonged bacterial residence. **(e)** Pie chart showing the proportion of total motile bacterial trajectories (12,241) classified into these four categories.

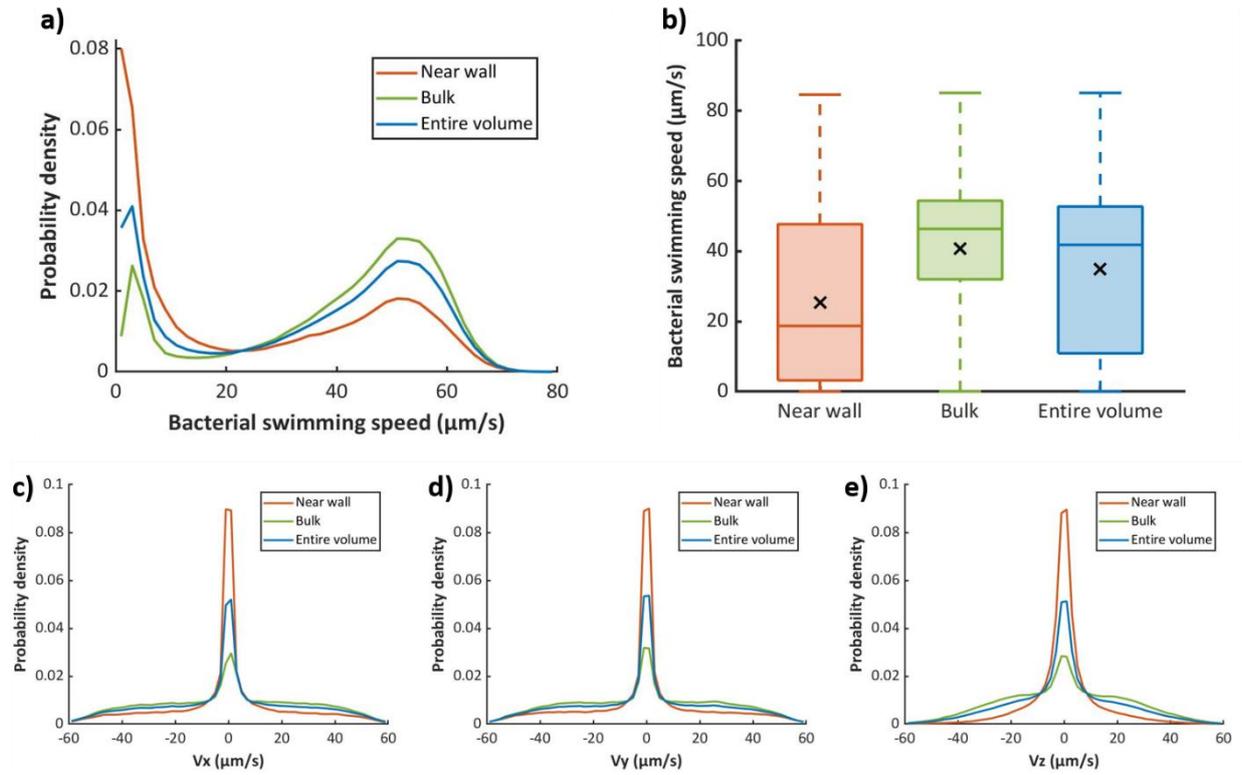

**Fig. S6. (a)** Bacterial swimming speeds distributions of motile *S. oneidensis* obtained in different regions of the capillary under static condition: near-wall regions ($z < 10$ µm and $z > 90$ µm), the bulk region ($z \in [0\,;90]$ µm) and the entire capillary volume. **(b)** Boxplots summarizing the swimming speed data for each region. The central line represents the median; boxes indicate the interquartile range, and the black cross represents the mean value. **(c–e)** Distributions of the velocity components associated with bacterial swimming speeds in the corresponding regions. Data were obtained from 6 independent experiments, with a total of 12,241 motile bacteria analyzed.

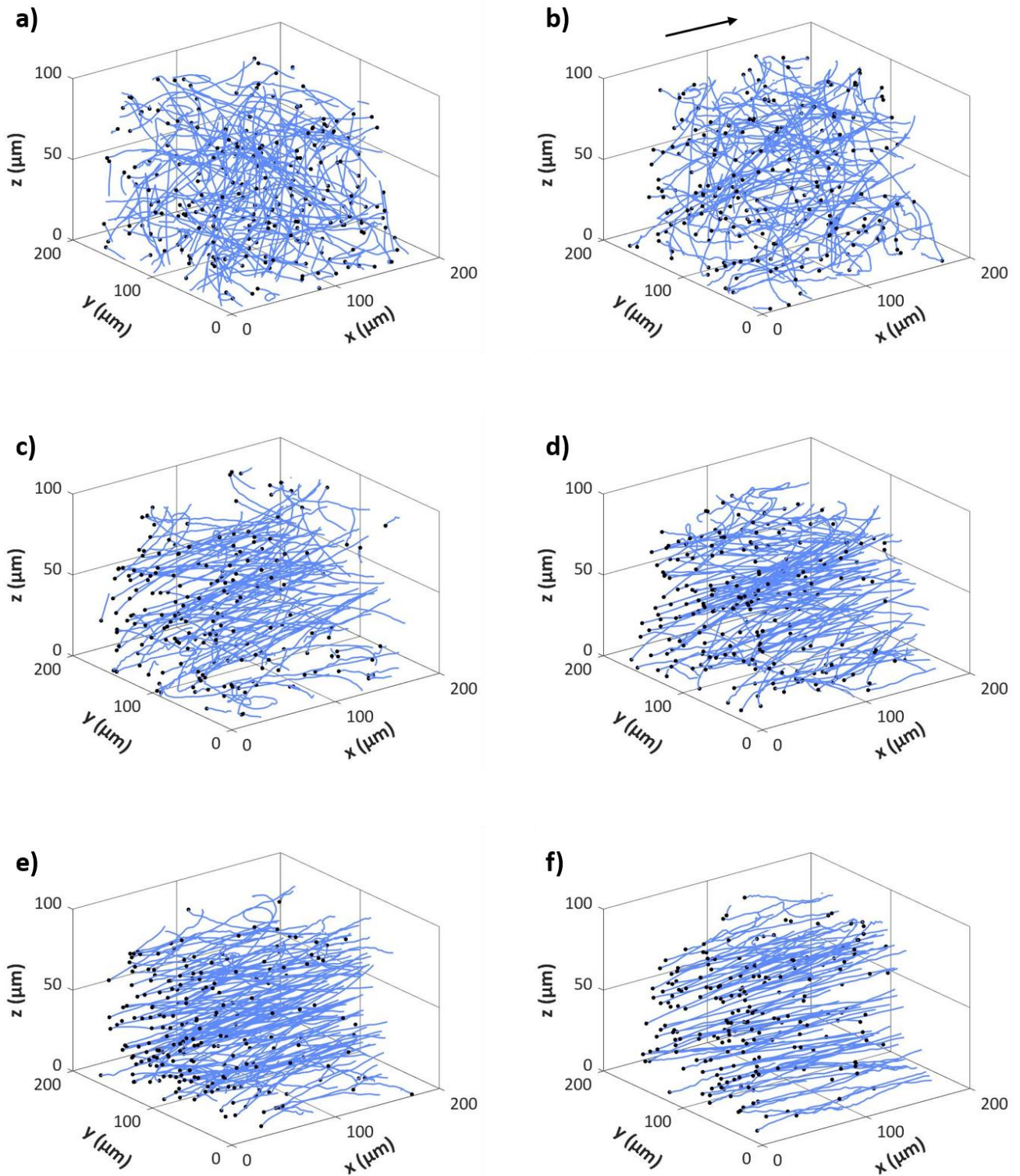

**Fig. S7.** Graph **(a)**, **(b)**, **(c)**, **(d)**, **(e)** and **(f)** are representative examples of three-dimensional trajectories (about 220) of motile wild-type *Shewanella oneidensis* obtained for different flow Peclet number (a = 0, b = 8, c = 19, d = 31, e = 38, f = 77) within a 180 × 180 × 100 µm³ volume. The capillary walls are located at z = 0 (bottom) and 100 µm (top). Flow direction is along the positive x-axis, as indicated by the black arrow in figure b.

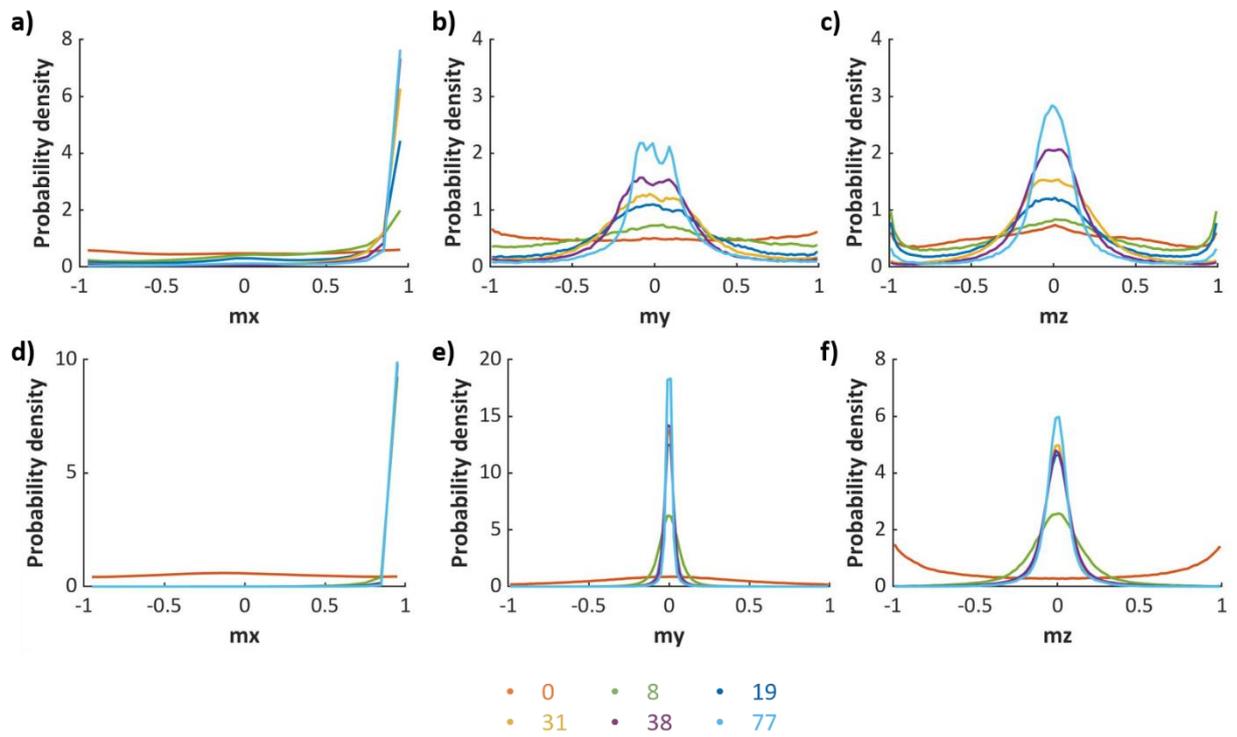

**Fig. S8.** Probability density of bacterial orientation in the x-direction **(a, d)**, the y-direction **(b, e)** and the z-direction **(c, f)** under applied flow Peclet number (0, 8, 19, 31, 38, 77) for the motile *S. oneidensis* WT population **(a, b, c)** and the Δfla mutant strain **(d, e, f)**. A total of 28,114 bacteria were analyzed for the motile WT population, and 84,103 for the Δfla mutant.

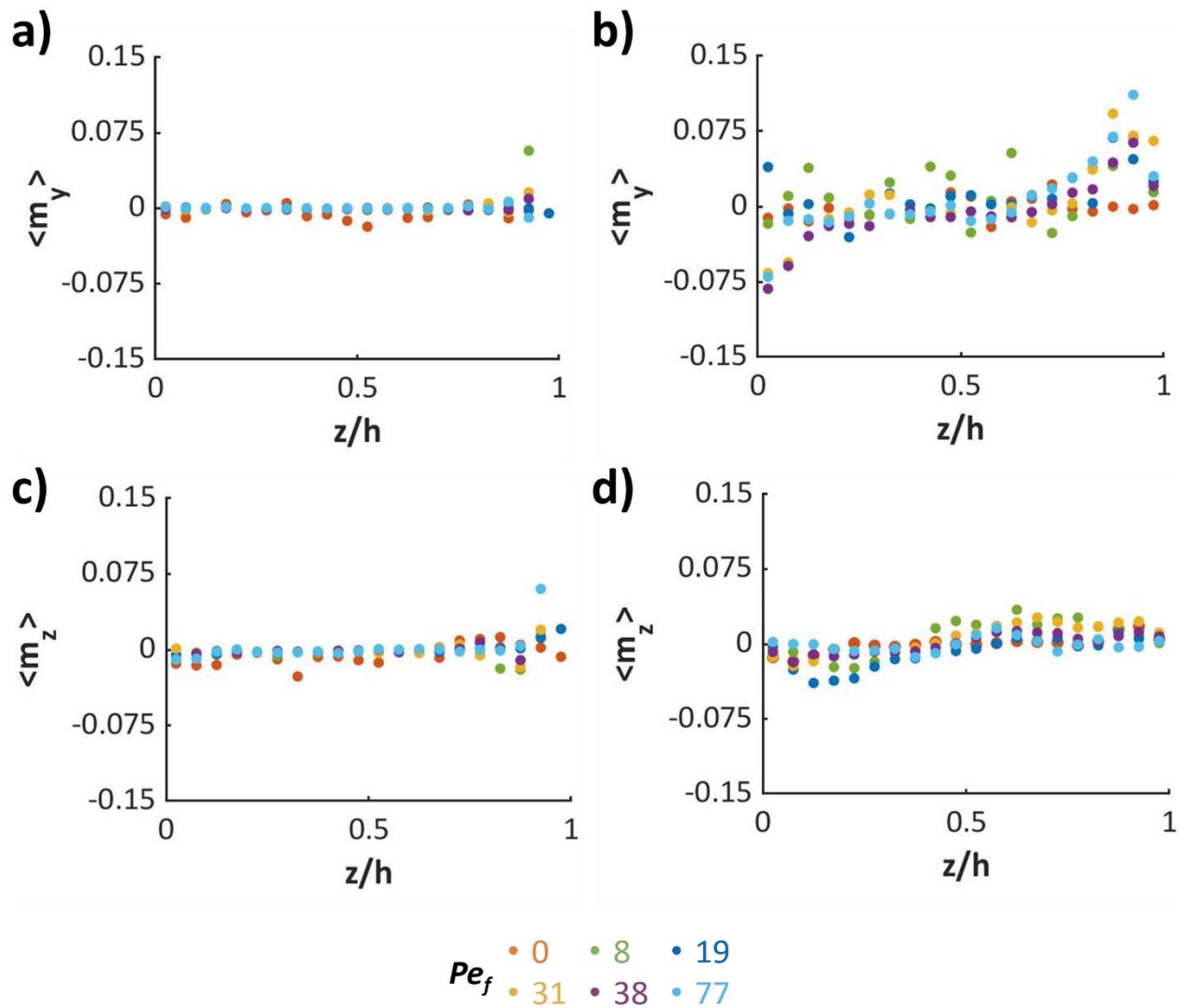

**Fig. S9. (a,b)** Plots showing the average y orientation ($m_y$) of the Δfla mutant **(a)** and the motile population of *S. oneidensis* WT **(b)** as a function of the flow Peclet number and their position within the capillary, normalized by its height (z/h). **(c,d)** Plots showing the average z orientation ($m_z$) of the Δfla mutant **(c)** and the motile population of *S. oneidensis* WT **(d)** as a function of the flow Peclet number and their position within the capillary, normalized by its height (z/h). z/h=0 corresponds to the bottom wall and z/h=1 corresponds to the top wall.
A total of 28,114 bacteria were analyzed for the motile WT population, and 84,103 for the Δfla mutant.

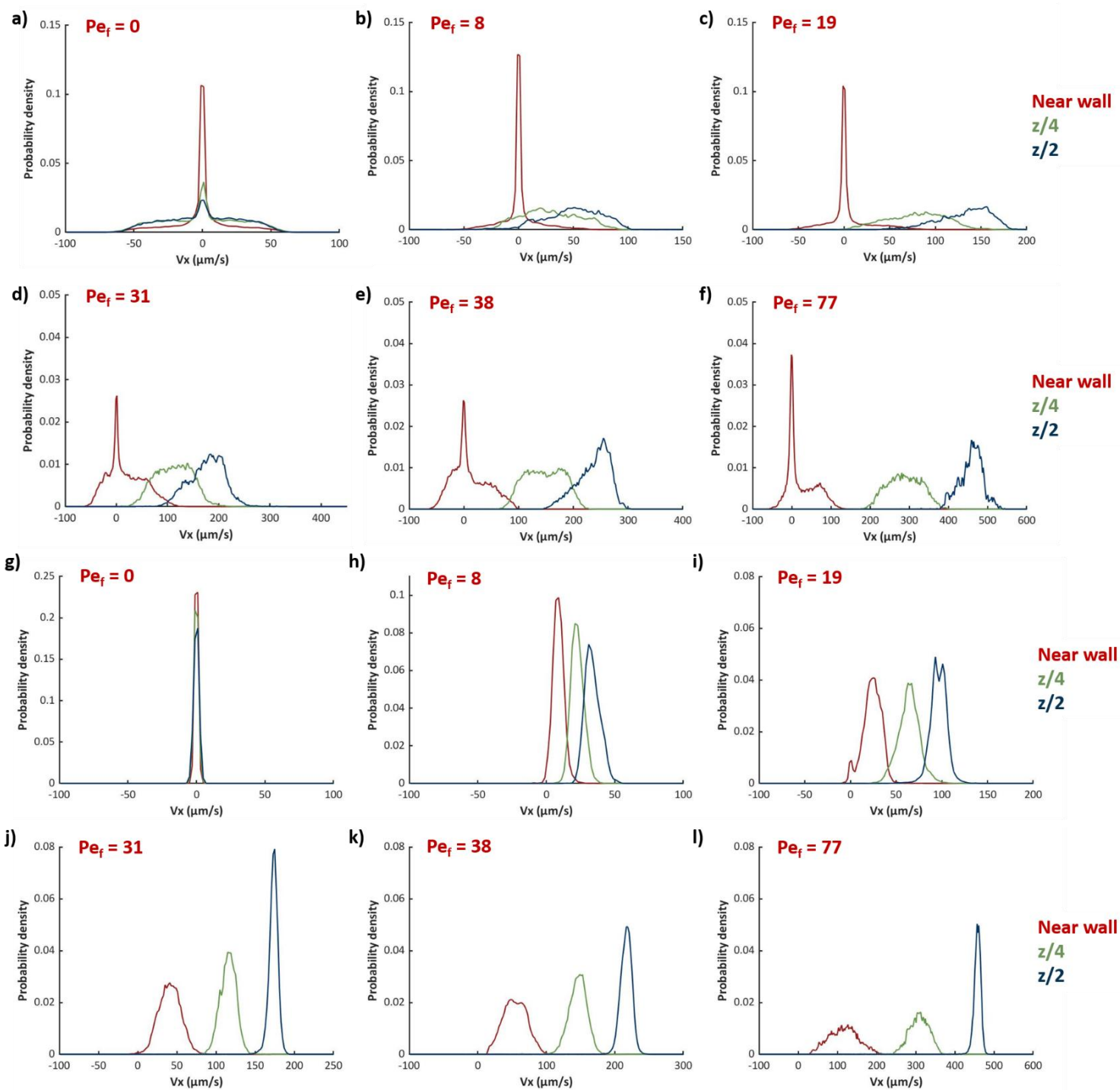

**Fig. S10.** Probability density of bacterial velocity component in the flow direction ($V_x$) for three distinct regions within the capillary: **(a)** near the walls (z < 5 μm and z > 95 μm) (dark red), **(b)** in the bottom and top quarter (z ϵ [20:25] and ϵ [75:80]) (green) and **(c)** at the central region (z ϵ [45:55]) (blue) of the capillary. Results are shown for motile *S. oneidensis* WT (a, b, c, d, e, f) and Δfla mutant (g, h, i, j, k, l) strains under applied flow Peclet number. A total of 28,114 bacteria were analyzed for the motile WT population, and 84,103 for the Δfla mutant.

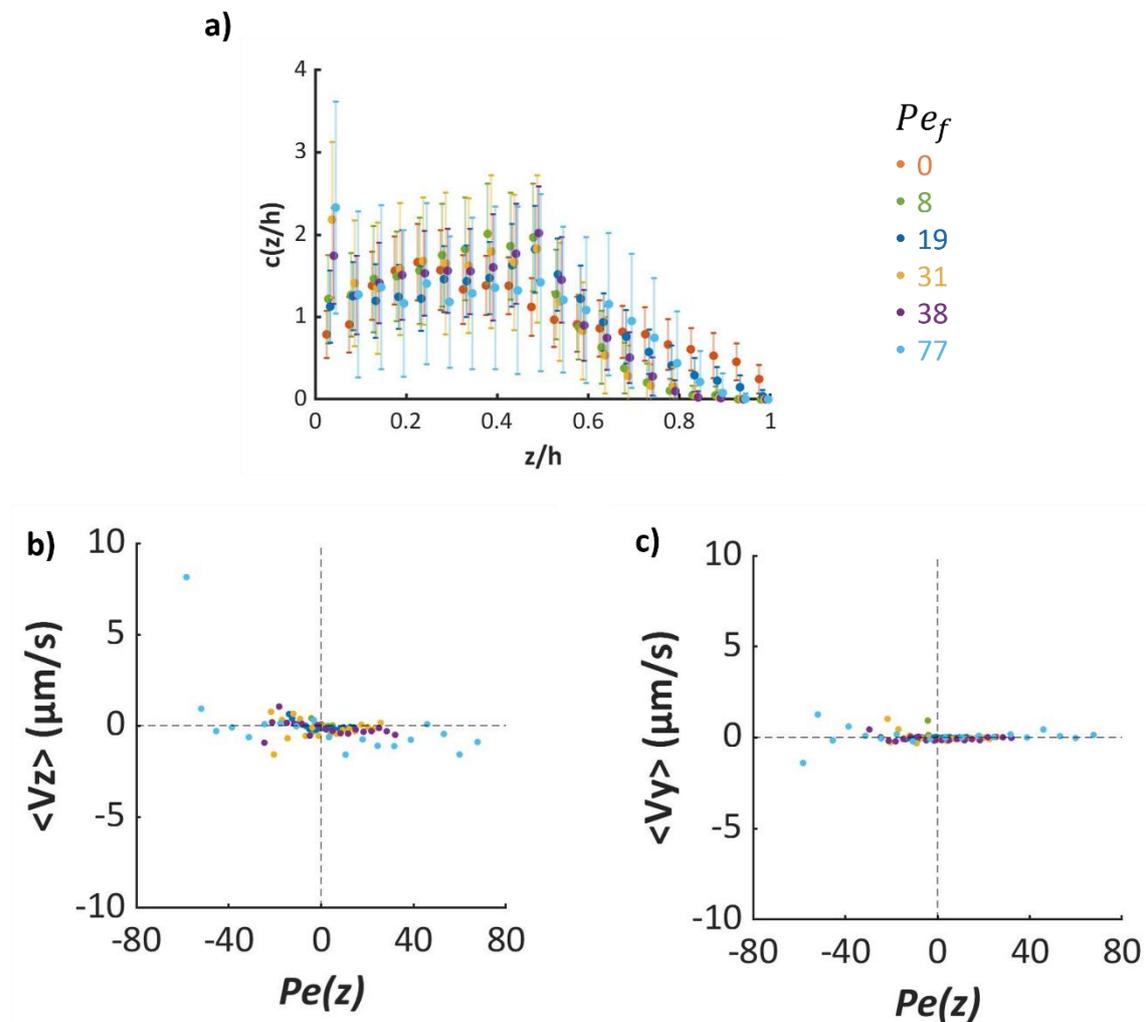

**Fig. S11. (a)** Vertical concentration profiles of *S. oneidensis* Δfla mutant from the bottom wall as a function of the normalized distance z across the capillary height. Curves are color-coded by the corresponding flow Peclet number. **(b-c)** Mean bacterial velocities: $V_z$ (μm/s) in panel (b) and $V_y$ (μm/s) in panel (c) as a function of the applied local Peclet number ($Pe_f$) for the *S. oneidensis* Δfla mutant. A total of 84,103 bacteria were analyzed.

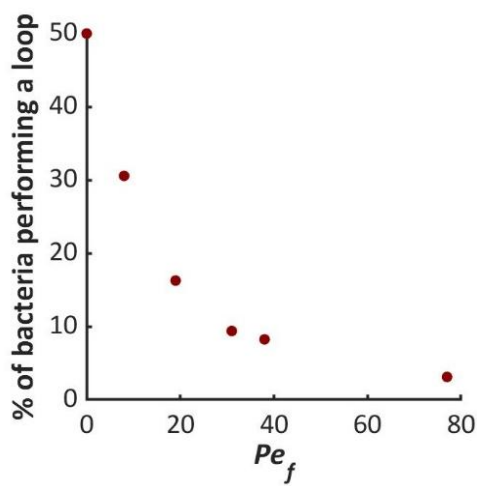

**Fig. S12:** Proportion of motile bacteria performing a loop during their trajectory over time, as a function of the applied local Peclet number ($Pe_f$). The total number of motile bacteria analyzed was 28,114.